\documentclass{article}

\usepackage{amsmath,amssymb,amsfonts}
\usepackage{verbatim}
\usepackage{graphicx}
\usepackage{color}
\begin{document}
	
	\title{{\bf Bargmann Representation of Quantum Absorption Refrigerators}}
	\author{ M.\ W.\ AlMasri$^{1}$ and M.\ R.\ B.\ Wahiddin$^{1,2}$\\	\small $^{1}$Cybersecurity and  Systems Research Unit, ISI-USIM,\\ \small Bandar Baru Nilai, 71800 Nilai, Negeri
		Sembilan, Malaysia.\\ \small $^{2}$ Department of Computer Science, Kulliyah of ICT, IIUM,\\ \small P.O. Box 10, 50728, Kuala Lumpur, Malaysia.}

	\maketitle
	
	\begin{abstract}In this work,  we solve the quantum absorption refrigerator analytically in the space of holomorphic functions with Gaussian measure . Our  approach simplifies the calculations since for a given quantum system the coordinate representation of any quantum state is always more complicated than its corresponding expression written with respect to the phase-space coordinate $z_{i}$.  We finally discuss the computational complexity of the holomorphic representation and compare it with the computational complexity of  the standard operator method and prove the efficiency of the holomorphic representation in computing some tasks. Our treatment is applicable to all quantum heat engines and refrigerators.  \vskip 5mm {\bf Keywords:} Bargmann representation, Quantum Absorption Refrigerators, Computational Complexity, Open Quantum Systems\end{abstract}
	
	\section{Introduction}Thermodynamics is one of the oldest and well-established branches of physics that sets boundaries to what can or cannot be achieved at  macroscopic level\cite{Landau}. It also has deep connection with other branches of science such as philosophy and computing theory\cite{Jos,david}. It was realized that large quantum devices, such as masers and  lasers, could be treated under the scope of  thermodynamics .  The pioneering work of Scovil et al. has shown the  equivalence of the Carnot engine  with  three-level Maser \cite{Wahiddin,scovil,scovil1}. This opened the road for flourishing  progress in the study of the connection between the laws of thermodynamics and quantum devices \cite{Alicki,Kosloff0,Geva,Nori,Mukamel,Kosloff,QT,Lutz}.\vskip 3mm

The interaction between any quantum mechanical system with environment may  lead to  dissipation or loss of information contained in the system to its environment.  
To obtain  a complete description of any quantum mechanical system, one needs to incorporate the effect of environment or baths to the original system Hamiltonian. This was the philosophy behind investigating open quantum systems \cite{Davies,Breuer,Rivas}. One of the main tasks in this theory is to solve the so-called Master equation of open quantum systems . Generally solving such equation can be very complicated and requires many tedious calculations especially in the non-Markovian case\cite{Huelga,Alonso}.\vskip 5mm In \cite{Fock}, Fock noticed the fact that one in principle   express the raising and lowering operators analytically in the complex $z$-plane as $z$ and $\frac{\partial }{\partial z}$ respectively. In this case the wavefunction should be written in term of the complex variable $z$ in a closed analytical form not as a vector in the Hilbert space. For example, the wavefunctions of the quantum  harmonic oscillator in coordinate representation are given by $\psi_{n}\left(x\right)= \frac{1}{\sqrt{2^{n} n!}} \left(\frac{m\omega}{\pi \hbar}\right)^{\frac{1}{4}} e^{-\frac{m\omega x^{2}}{2\hbar}} H_{n}\left(\sqrt{\frac{m \omega }{\hbar}}x\right) $ where $H_{n}\left(y\right)= \left(-1\right)^{n} e^{y^{2}} \frac{d^{n}}{dy^{n}} \left(e^{-y^{2}}\right)$ is the Hermite polynomials while in the Bargmann representation it  is simply  $\psi_{n}\left(z\right)=\frac{z^{n}}{\sqrt{n!}}$ written with respect to the phase space coordinate $z$ .  The rigorous treatment of the holomorphic function spaces described here was done by Segal and Bargmann in 60s. They also provided a transform formula where one could obtain the coordinate representation of wavefunctions from the monomials $\psi_{n}\left(z\right)=\frac{z^{n}}{\sqrt{n!}}$ \cite{Bargmann,Segal}. The Bargmann representation of some quantum mechanical system has been studied before such as for Jaynes-Cummings model, thermal coherent states and WKB approximation\cite{Stenholm,Bishop,Voros}. Very interestingly it was shown that  Bargmann representation gives better results in the case of WKB approximation  comparing with other known methods\cite{Voros}. In \cite{Thiemann}, a generalization of the Segal-Bargmann transform has been used in the context of canonical gravity and loop quantum gravity. 
\\  For sake of simplicity  we systematically discuss  the theory of open quantum systems in term of  holomorphic functions.  Moreover we solve the  quantum absorption refrigerators driven by a Gaussian noise as a model example  in term of analytical functions in the complex $z$-plane \cite{levy,Potts}. Recently the quantum absorption refrigerator  was  realized experimentally in the case of trapped ions \cite{Maslennikov}. 

\section{Bargmann representation of open quantum systems}

{\bf Definition 1:}	The Bargmann space also known as the Fock-Bargmann or Segal-Bargmann space, denoted by $\mathcal{H}L^{2}\left(\mathbb{C}^{n}, \mu\right)$, is the space of holomorphic functions with $\mu\left(z\right)= ( \pi)^{-n} e^{-|z|^{2}}$ and  $|z|^{2}= |z_{1}|^{2}+\dots+ |z_{n}|^{2}$. Any entire analytic  function $f\left(z\right)$ in this space  obeys   the following  square-integrability condition \cite{Bargmann,Segal,Askold,Folland,Hall1,Ali,Hall} 
\begin{equation}\label{condition}
||f||^{2}: =\langle f| f\rangle_{\mu}= \left(\pi\right)^{-n}\int_{\mathbb{C}^{n}} |f\left(z\right)|^{2} e^{-|z|^{2}}dz < \infty, 
\end{equation}
where  $dz$ is the $2n$-dimensional Lebesgue measure on $\mathbb{C}^{n}$. \vskip 5mm
{\it Remark 1:} The inner-product of any two analytic functions $f\left(z\right)$ and $g\left(z\right)$ satisfying the condition \ref{condition} is 
\begin{equation}\label{inner}
\langle f|g\rangle_{\mu}= \left(\pi\right)^{-n}\int_{\mathbb{C}^{n}}	 \overline{f}\left(z\right) g\left(z\right) e^{-|z|^{2} } dz. 
\end{equation}

Using the inner product defined in \ref{inner}, we can prove that both  $\overline{z}$ and $\frac{\partial}{\partial z}$ have the same effect i.e. $\langle\frac{\partial f}{\partial z}, g\rangle_{\mu} = \langle f,z g\rangle_{\mu}$ \cite{Hall1}. 
\vskip 5mm
{\bf Lemma 1:}	{\it  The Bargmann space $\mathcal{H}L^{2}\left(\mathbb{C}^{n}, \mu\right)$ is  a Hilbert space}. \\ This observation was due to Bargmann in \cite{Bargmann}. More generally, $L^{2}$ is the only Hilbert space among the function spaces $L^{p}$\cite{Folland2}.\\

According to  Riesz representation theorem between  any given Hilbert space and its dual space there must exist a unique $f_{\alpha}$ such that 
\begin{equation}
f\left(\alpha\right)=\langle f_{\alpha}\left(z\right)|f\left(z\right)\rangle_{\mu}, 
\end{equation}
where the function $f_{\alpha}$ is usually called a coherent state with parameter $\alpha$ whereas the quantity $\overline{f_{\alpha}}$ is known as the reproducing kernel \cite{Aronszajn,Bergman}. \vskip 5mm
{\bf Lemma 2:}
{\it 	The reproducing kernel in the Bargmann space is } 
\begin{equation}
K\left(z,w\right)= \sum_{n=0}^{\infty} \frac{z^{n}}{\sqrt{n!}}\frac{\overline{w}^{n}}{\sqrt{n!}}= \sum^{\infty}_{n=0} \frac{1}{n!}\left(z\overline{w}\right)^{n} = e^{z\overline{w}}. 
\end{equation}
\vskip 5mm
{\bf Lemma 3:} {\it	The monomials $z^{n}/\sqrt{n!}$ or its alternate expression $z^{n}/\sqrt{\Gamma\left(n+1\right)}$ form an  orthonormal  basis }
\begin{equation}
\int \frac{dz\;  d\overline{z}}{\pi } \exp[-z \overline{z}] \; \overline{z}^{n} z^{m} = n!\; \delta_{m n}
\end{equation}
Thus the  bosonic wave functions can be written as a uniformly convergent  series $f\left(z\right)= \sum_{n} c_{n} \frac{z^{n}}{\sqrt{n!}}$. The convergence of this series in any compact domain of the complex $z$-plane is fixed by the condition $\sum^{\infty}_{n} |c_{n}|^{2}=1$ \cite{Askold,Boas}. 

\vskip 5mm

{\bf Lemma 4:} {\it 	In Bargmann representation, the raising and lowering  operators can be defined  as $a^{\dagger}_{i}= z_{i}$ and  $a_{i}= \frac{\partial}{\partial z_{i}} $ or $\overline{z}_{i}$}.    \\
This can be proven by straightforward calculations of the commutators, 
\begin{align}
\left[\frac{\partial}{\partial z_{i}},z_{j}\right]= \delta_{i j}, \\ 
\left[\frac{\partial}{\partial z_{i}},\frac{\partial}{\partial z_{j}}\right]=\left[z_{i},z_{j}\right]=\left[\overline{z}_{i},\overline{z}_{j}\right]=0.
\end{align}	
\vskip 5mm
With this construction we define the position and momentum operators in natural units i.e. $\hbar=1$ as 
\begin{eqnarray}
x_{j}= \frac{1}{2}\left(\frac{\partial}{\partial z_{j}}+ z_{j}\right), \\ 
p_{j}= \frac{1}{2i}\left(\frac{\partial}{\partial z_{j}}- z_{j}\right).
\end{eqnarray}
Since the operators $x_{j}$ and $p_{j}$ satisfy the Weyl relations and act  irreducibly  in the Bargmann space, one could in principle as consequence  of the Stone-von-Neumann  theorem maps each position dependent quantity in the Hilbert space $L^{2}\left(\mathbb{R}^{2n}\right)$ to its corresponding  holomorphic expression  in the Bargmann space $\mathcal{H}L^{2}\left(\mathbb{C}^{n}, \mu\right)$\cite{Hall}. This can be done in virtue of 
the Segal-Bargmann transform  given by \cite{Ali,Hall}
\begin{equation}
\left(Af\right)\left(z\right)=\int_{\mathbb{R}^{n}} \exp\left[-\left(z\cdot z-2\sqrt{2}z\cdot x+ x\cdot x\right)\right]f\left(x\right)\;  dx, 
\end{equation}   \vskip 5mm
As an example the harmonic oscillator Hamiltonian (up to a constant term) $H_{0}=\hbar \omega a^{\dagger} a$ assumes the following form in the Bargmann representation $H_{0}=  
\hbar \omega  z \frac{d}{dz}$ \cite{Zinn-Justin}. Acting by $H_{0}$ to the energy eigenstates we get ,
\begin{equation}
H_{0}|n\rangle= \hbar \omega z \frac{d}{dz} \frac{z^{n}}{\sqrt{n!}} = n \hbar \omega \frac{z^{n}}{\sqrt{n!}}= n \hbar \omega |n\rangle.
\end{equation} 	
\vskip 5mm

Another interesting example from  quantum optics is the  cat-states which are defined as the quantum superposition of two opposite-phase coherent states of a single mode\cite{Wahiddin,cats,Glauber}. 
The even and odd cat states can be defined respectively as \cite{Dodonov}
\begin{align}\label{states}
|\alpha_{+}\rangle= |\alpha\rangle+ |-\alpha\rangle= e^{-\frac{1}{2}|\alpha|^{2}} \sum_{n=0}^{\infty} \frac{\alpha^{2n}}{\sqrt{2n!}} |2n\rangle, \\  \label{states1}
|\alpha_{-}\rangle=|\alpha\rangle- |-\alpha\rangle= e^{-\frac{1}{2}|\alpha|^{2}} \sum_{n=0}^{\infty} \frac{\alpha^{2n+1}}{\sqrt{\left(2n+1\right)!}} |2n+1\rangle, 
\end{align}
where $|\alpha\rangle= e^{-\frac{1}{2}|\alpha|^{2}} \sum_{n=0}^{\infty} \frac{\alpha^{n}}{\sqrt{n!}} |n\rangle= e^{-\frac{1}{2}|\alpha|^{2}}\sum_{n=0}^{\infty} \frac{\alpha^{n} \left(a^{\dagger}\right)^{n}}{n!} |0\rangle $ in the Bosonic Fock number basis \cite{Glauber}.

The even and odd cat states defined in \ref{states} and \ref{states1} can be rewritten in a closed analytic form using the Bargmann representation as 
\begin{align}\label{analytic1}
|\alpha_{+}\rangle= |\alpha\rangle+ |-\alpha\rangle=2 e^{-\frac{1}{2}|\alpha|^{2}} \sum_{n=0}^{\infty} \frac{\left(\alpha z\right)^{2n}}{2n!} \\ \nonumber =2 e^{-\frac{1}{2}|\alpha|^{2}} \cosh\left(\alpha\right) z= \sqrt{2\pi \alpha z }\;  e^{-\frac{1}{2}|\alpha|^{2}} I_{-1/2}\left(\alpha z\right) , \\ \label{analytic2}
|\alpha_{-}\rangle=|\alpha\rangle- |-\alpha\rangle= 2e^{-\frac{1}{2}|\alpha|^{2}} \sum_{n=0}^{\infty} \frac{\left(\alpha z\right)^{2n+1}}{\left(2n+1\right)!}\\ \nonumber = 2 e^{-\frac{1}{2}|\alpha|^{2}} \sinh\left(\alpha z\right)= \sqrt{2\pi \alpha z }\;  e^{-\frac{1}{2}|\alpha|^{2}} I_{1/2}\left
(\alpha z\right), 
\end{align}
since 
\begin{equation}
|\alpha\rangle= e^{-\frac{1}{2}|\alpha|^{2}} \sum_{n=0}^{\infty} \frac{\alpha^{n}}{\sqrt{n!}} |n\rangle= e^{-\frac{1}{2}|\alpha|^{2}}\sum_{n=0}^{\infty} \frac{\left(\alpha z\right)^{n}}{n!} =  e^{-\frac{1}{2}|\alpha|^{2}} e^{\alpha z}, 
\end{equation}
and $I_{1/2,-1/2}$ are the modified Bessel functions.
Interestingly the equations \ref{analytic1} and \ref{analytic2} are written with respect to complex variables  $\alpha$ and $z$ . As far as we are aware, these equations has not been given explicitly before in the literature . 
It would be interesting to see if they will reproduce all physical results found previously for the even and odd cat states analytically.\vskip 5mm

In  case of fermions, we can construct analytical  representation on the fermionic phase space $\mathbb{R}^{2d}$ by realizing the raising and lowering operators in term of the Grassmann variable $\theta$ as 
\begin{eqnarray}
b_{j}= \frac{\partial}{\partial \theta_{j}}, \\
b^{\dagger}_{j}= \theta_{j},
\end{eqnarray}
where $\theta_{j}= \frac{1}{2}\left(\xi_{2j-1}- i \xi_{2j}\right)$, $\overline{\theta}_{j}=  \frac{1}{2} \left(\xi_{2j-1}+i \xi_{2j}\right) $ with $j=1,\dots d$ , $\left[\xi_{i},\xi_{k}\right]= 2 i \epsilon_{ikl} \xi_{k}$ and $\left\{\xi_{i},\xi_{k}\right\}=0$ for $i\neq k$ and 2 for $i=k$. It is important to note that this specific construction is when the fermionic phase space is for  spin-1/2 particles or in general any two-level systems  and this explains the 2 in front of the commutation relation between any two different  components of the vector $\xi$.  For higher spins the generalization is not  difficult and one only need to generalize the Grassmann algebra for higher spins and write the exact commutation and anti-commutation relations for each case . The fermionic phase space is even since the complex structure should be preserved.   With this construction, the anticommutation relations are 
\begin{equation}\label{anti}
\left\{\theta_{i},\theta_{j}\right\}= \left\{\overline{\theta}_{i}, \overline{\theta}_{j}\right\}= 0 , \quad \left\{\overline{\theta}_{i}, \theta_{j}\right\}= \delta_{ij}. 
\end{equation}
The Gaussian integral over complex Grassmann (anti-commuting) variables is 
\begin{equation}
\int d\overline{\theta} d\theta\;  e^{-\overline{\theta} b \theta}= b , 
\end{equation}
From this result we find
\begin{equation}
\int d\overline{\theta} d\theta \; \theta \overline{\theta}\; e^{-\overline{\theta} b \theta}= 1. 
\end{equation} 
The basic integral  rules for a single Grassmann variable $\theta$ are given by the  Berezin integrals defined as \cite{Cartier,Takhtajan}
\begin{eqnarray}
\int \theta d\theta=1, \; \; \; \int d\theta=1 . 
\end{eqnarray}
Applying  Berezin integrals over the most general one-variable  Grassmann polynomial $f\left(\theta\right)= a\theta+b$ gives 
\begin{equation}
\int  f\left(\theta\right) d\theta= \int \left(a\theta+b\right)d\theta = a , 
\end{equation}
where $a$ and $b$ are elements from  $\mathbb{C}$. 
The orthogonal basis for a single fermion are simply ${\bf 1}=|0\rangle$ and $\theta= |1\rangle$ since $\theta^{2}=0$.
\\ 

The total fermionic number operator for many-body fermionic system  is 
\begin{equation}
N_{F}= \sum_{i}\theta_{i}\; \overline{\theta}_{i}, 
\end{equation}
and  satisfies  $	N^{2}_{F_{i}}= 	N_{F_{i}}$. \\  By direct calculation we find for each $i$
\begin{equation}
N^{2}_{F_{i}}= \theta_{i}\overline{\theta}_{i} \theta_{i}\overline{\theta}_{i} =\theta_{i} \left(1-\theta_{i}\overline{\theta}_{i}\right)\overline{\theta_{i}}= \theta_{i}\overline{\theta}_{i}=N_{F_{i}}
\end{equation}
since according to  \ref{anti}, we have   $\theta_{i}^{2}=\overline{\theta}_{i}^{2}=0$.
\vskip 5mm
The total Hamiltonian should  be written as a function of $z_{i}$, $\theta_{i}$ and their complex conjugates  i.e. $H\left(z_{i},\overline{z}_{i}, \theta_{i}, \overline{\theta}_{i}\right)$ in general.  Moreover the corresponding bosonic eigenstates are written as a polynomial of complex variable $z_{i}$ 
\begin{equation}
|n_{1}\dots , n_{k}\rangle= \prod_{i=1}^{i=k} \frac{ z^{n_{k}}}{\sqrt{n_{k}!}}, 
\end{equation} 
and the fermionic eigenstates for each single fermion  are 1 for the ground state and $\theta$ for the excited state. 
\vskip 5mm 

As a model example we consider the Jaynes-Cummings model that describes the interaction between two-level atom and quantized field in an optical cavity.  The Jaynes-Cummings Hamiltonian  up to a constant term in operator language reads  \cite{Wahiddin,Vogel}
\begin{equation}
\hat{H}_{\mathrm{JC}}= \hbar \omega_{c} \hat{a}^{\dagger} \hat{a} + \hbar \omega_{eg} |e\rangle \langle e| + \hbar g_{c}\left(\hat{\sigma}_{+}\hat{a}+ \hat{\sigma}_{-}\hat{a}^{\dagger}\right), 
\end{equation}
where $\omega_{c}$ is the cavity frequency and $\omega_{eg}$ is the resonance frequency of the transition between atomic sub-levels. $\hat{a},\hat{a}^{\dagger}$ are the bosonic raising and lowering operators of the cavity, $\hat{\sigma}_{+}= |e \rangle \langle g|, \hat{\sigma}_{-}= |g\rangle \langle e|$ are the raising and lowering operators of the atom. Moreover we can define the atomic inversion operator as $\hat{\sigma}_{z}= |e\rangle \langle e|- |g \rangle \langle g|$. We  re-write the analytical version  of the Jaynes-Cummings Hamiltonian as
\begin{equation}
H_{\mathrm{JC}}= \hbar \omega_{c} z \overline{z} + \hbar \omega_{eg} \theta \overline{\theta}+ \hbar g_{c} \left(\theta \overline{z}+ \overline{\theta} z\right), 
\end{equation}
or equivalently 
\begin{equation}
H_{\mathrm{JC}}= \hbar \omega_{c} z \frac{\partial }{\partial z} + \hbar \omega_{eg} \theta \frac{\partial }{\partial \theta}+ \hbar g_{c} \left(\theta \frac{\partial }{\partial z}+ \frac{\partial }{\partial \theta} z\right), 
\end{equation}
where we identified  the excited and ground states of the atom as $|e\rangle=\theta$ and $|g\rangle=1$. We many define the number operator as 
\begin{equation}
N= \theta \overline{\theta}+ z \overline{z}= \theta \frac{\partial}{\partial \theta}+ z \frac{\partial }{\partial z}. 
\end{equation}
The eigenstates of the  number operator  commutes with the atom-field Hamiltonian $[N,H_{\mathrm{JC}}]=0$. Thus it can be used as a basis of the tensor product states $|e,n\rangle, |g,n\rangle, |e,n-1\rangle \dots $. However in the analytical formulation  we express these states as a product of monomials not as a tensor product of state vectors. For example, the matrix element 
\begin{eqnarray}
\langle g,n|\hat{H}_{\mathrm{JC}} | e,n-1\rangle = \hbar g_{c} \langle g,n| \hat{a}^{\dagger } \hat{\sigma}_{-}| e,n-1\rangle + \hbar g_{c} \langle g,n| \hat{a} \hat{\sigma}_{+} | e, n-1\rangle = \hbar \sqrt{n} g_{c}, 
\end{eqnarray}
can be calculated analytically as 

\begin{align}\label{proof}
\langle g,n|H_{\mathrm{JC}} | e,n-1\rangle= \hbar g_{c} \langle g,n| z \overline{\theta}| e,n-1\rangle + \hbar g_{c} \langle g,n| \overline{z} \theta | e, n-1\rangle\\ 
= \hbar g_{c} \langle g| \overline{\theta}|e\rangle \langle n| z| n-1\rangle_{\mu}+ \hbar g_{c} \langle g|\theta|e \rangle \langle n, \overline{z}|n-1\rangle_{\mu}\\
= \hbar g_{c}  \langle g| \overline{\theta}|e\rangle \langle n| z| n-1\rangle_{\mu}= \hbar g_{c}  \langle g| \overline{\theta}|e\rangle \int \frac{dz d\overline{z}}{\pi}\frac{\overline{z}^{n}}{\sqrt{n!}} e^{-|z|^{2} } z \frac{z^{n-1}}{\sqrt{\left(n-1\right)!}}\\ 
= \hbar g_{c} \int \frac{dz d\overline{z}}{\pi}\frac{\overline{z}^{n}}{\sqrt{n!}} e^{-|z|^{2} }  \frac{z^{n}}{\sqrt{\left(n-1\right)!}}\\
= \sqrt{n} \hbar g_{c}\int \frac{dz d\overline{z}}{\pi}\frac{\overline{z}^{n}}{\sqrt{n!}} e^{-|z|^{2} } \frac{z^{n}}{\sqrt{n}!}= \sqrt{n} \hbar g_{c}, 
\end{align}

since $|e\rangle= \theta$, thus $\theta|e\rangle= \theta^{2}=0$, also $\frac{\partial }{\partial z}|n-1\rangle=\sqrt{n-1} |n-2\rangle$ thus the quantity $ \langle g,n| \overline{z} \theta | e, n-1\rangle$=  $\langle g,n| \theta \frac{\partial }{\partial z} | e, n-1\rangle= 0$ while $\langle g| \overline{\theta}|e\rangle=1$ since $\overline{\theta}|e\rangle= \overline{\theta} \theta={\bf 1}=|g\rangle$ and $\langle g|g\rangle=1$ in an orthonormal basis. The previous calculations show that one can separate the cavity states from the atomic level states safely and express them as a product of analytical quantities so no need to worry about the matrix dimensionality of the constructed space from tensor product of the cavity and atomic states and other related issues.   Finally we can express the eigenstates of Jaynes-Cummings Hamiltonian as 
\begin{eqnarray}
|n, +\rangle= \cos \left(\frac{\phi_{n}}{2}\right)|e, n-1\rangle+ \sin\left(\frac{\phi_{n}}{2}\right)|g,n\rangle\\ \nonumber
= \cos \left(\frac{\phi_{n}}{2}\right) \theta \frac{z^{n-1}}{\sqrt{\left(n-1\right)!}}+ \sin\left(\frac{\phi_{n}}{2}\right) {\bf 1} \frac{z^{n}}{\sqrt{n!}}, \\
|n,-\rangle= \cos \left(\frac{\phi_{n}}{2}\right) |g,n\rangle-\sin\left(\frac{\phi_{n}}{2}\right) |e,n-1\rangle\\ 
\nonumber = \cos \left(\frac{\phi_{n}}{2}\right)  {\bf 1} \frac{z^{n}}{\sqrt{n!}} - \sin\left(\frac{\phi_{n}}{2}\right) \theta \frac{z^{n-1}}{\sqrt{\left(n-1\right)!}}, 
\end{eqnarray}
where ${\bf 1}$ is the fermionic ground state and obeys the Grassmann algebra rules. \vskip 5mm
For a given  open quantum systems, we  connect the system with one or more reservoirs (baths). In the case of system coupled to a single bath we have the Hamiltonian  \cite{Davies}
\begin{equation}
H= H_{\mathrm{s}}+ H_{\mathrm{b}}+ H_{\mathrm{sb}}, 
\end{equation}
where $H_{\mathrm{s}}$ is the system Hamiltonian, $H_{\mathrm{b}}$ is the bath Hamiltonian and $H_{\mathrm{sb}}$ is the system-bath interaction term. In the Markovian approximation in which
the time derivative of the operator depends on the operator itself not its histories, the evolution of any observer $O$ such as the density matrix $\rho_{\mathrm{sb}}$ is given by the Gorini–Kossakowski–
Sudarshan–Lindblad (GKS-L) equation \cite{Kossakowski,Gorini,Spohn1}
\begin{equation}
\frac{d}{dt}O= -\frac{i}{\hbar} \left[H_{\mathrm{s}},O\right]+ \mathcal{L}_{D}\left(O\right), 
\end{equation}
where	$\mathcal{L}_{D}$ is the  dissipative  part or dissipator  given by \cite{Lindblad}
\begin{equation}
\mathcal{L}_{D}\left(O\right)= \sum_{n} \gamma_{n} \left(L_{n}O\overline{L}_{n}-\frac{1}{2}\left\{\overline{L}_{n}L_{n},O\right\}\right), \quad  \gamma_{n}\geq0, 
\end{equation}
and  $L_{n}$ are  the  Lindblad jump operators and $\overline{L}_{n}$ are their complex conjugates. 
\\ Let $\{\Lambda_{s}|s\geq 0\}$ be a quantum dynamical semigroup and let $\rho^{0}\in T\left(\mathcal{H}L^{2}\left(\mathbb{C}^{n}, \mu\right)\right)$ be an $\Lambda_{s}$-invariant state where $T\left(\mathcal{H}L^{2}\left(\mathbb{C}^{n}, \mu\right)\right)$ is the Banach space of trace class operators, then    the associated entropy production $\sigma$  relative to $\rho^{0}$  is \cite{Spohn}
\begin{equation}
\sigma(\rho)= -\frac{d}{dt} S\left(\Lambda_{s} \rho|\rho^{0}\right)|_{s=0}. 
\end{equation}
As a final comment in this section, the holomorphic representation works also in the non-Markovian approximation and in any quantum mechanical system indeed. 

\section{Bargmann representation solution of quantum absorption refrigerator }
The Hamiltonian of quantum absorption refrigerator  is composed of three  interacting oscillators \cite{Kosloff}
\begin{align}\label{main}
H_{\mathrm{s}}= H_{0}+ H_{\mathrm{int}}, \\
H_{0}= \hbar \omega_{\mathrm{h}} z_{\mathrm{h}}\frac{\partial}{\partial z_{\mathrm{h}}}+\hbar  \omega_{\mathrm{c}} z_{c}\frac{\partial}{\partial z_{c}}+\hbar  \omega_{\mathrm{w}} z_{\mathrm{w}}\frac{\partial}{\partial z_{\mathrm{w}}}=\sum_{\mathrm{i}}\hbar \omega_{\mathrm{i}} z_{\mathrm{i}}\frac{\partial}{\partial z_{\mathrm{i}}}, \\
H_{\mathrm{int}}= \hbar \omega_{\mathrm{int}} \left(z_{\mathrm{h}} \frac{\partial}{\partial z_{\mathrm{c}}}\frac{\partial}{\partial z_{\mathrm{w}}}+ z_{\mathrm{c}}z_{\mathrm{w}} \frac{\partial}{\partial z_{\mathrm{h}}}\right), 
\end{align}
where $\mathrm{i}=\mathrm{h},\mathrm{c},\mathrm{w}$ denotes the hot,cold and work reservoirs. 
The unperturbed energy eigenstates for the quantum absorption refrigerator  in the Bargmann representation are $|n_{\mathrm{h}},n_{\mathrm{c}},n_{\mathrm{w}}\rangle^{(0)}=  \frac{1}{\sqrt{n_{\mathrm{h}}!n_{\mathrm{c}}!n_{\mathrm{w}}!}} z^{n_{\mathrm{h}}}_{\mathrm{h}}z^{n_{\mathrm{c}}}_{\mathrm{c}}z^{n_{\mathrm{w}}}_{\mathrm{w}}$.
Under steady-state conditions, the first and second laws of thermodynamics should be satisfied, namely the following relations holds
\begin{align}
\mathcal{J}_{\mathrm{h}}+ \mathcal{J}_{\mathrm{c}}+ \mathcal{P}=0, \\
-\frac{\mathcal{J}_{\mathrm{h}}}{T_{\mathrm{h}}}-\frac{\mathcal{J}_{\mathrm{c}}}{T_{\mathrm{c}}}-\frac{\mathcal{P}}{T_{\mathrm{w}}}\geq 0,
\end{align}
where the first relation represents the conservation of energy( first law of thermodynamics) while the second inequality states that the sum of entropies is equal or greater than 0(second law of thermodynamics, positive production of entropy in the  universe).\\
Quantum mechanically, we may re-write the Hamiltonian $\ref{main}$ as 
\begin{equation}
H=H_{0}+\lambda H_{\mathrm{int}}. 
\end{equation}
Assuming the energy spectrum to be non-degenerate, we apply the holomorphic perturbation theory where we expand the energy eigenstates in term of phase-space coordinates $\left\{z_{\mathrm{i}}\right\}$ not $\left\{\psi(x_{\mathrm{i}})\right\}$. The energy levels and eigenstates of the perturbed Hamiltonian are given by time-independent  Schrödinger equation
\begin{equation}
(H_{0}+\lambda H_{\mathrm{int}})|n\rangle= E_{n}|n\rangle . 
\end{equation}
If we assume the perturbation to be extremely small, we may expand the energy levels and eigenstates as power series 
\begin{eqnarray}
E_{n}|n\rangle = E^{(0)}_{n}+\lambda E^{(1)}_{n}+ \lambda^{2}E^{(2)}_{n}+\dots \\
|n\rangle= |n^{(0)}\rangle+\lambda|n^{(1)}\rangle + \lambda^{2}|n^{(2)}\rangle+\dots
\end{eqnarray} 
The first-order energy shift in the quantum absorption refrigerator  is 
\begin{eqnarray}
E^{(1)}_{n}=\langle m^{(0)}|H_{\mathrm{int}}|n^{(0)}\rangle_{\mu}\\ 
=\hbar \omega_{\mathrm{int}}\big(\;  \delta_{m_{\mathrm{h}},n_{\mathrm{h}+1}}\delta_{m_{\mathrm{c}},n_{\mathrm{c}-1}} \delta_{m_{\mathrm{w}},n_{\mathrm{w}-1}}\sqrt{(n_{\mathrm{h}}+1)n_{\mathrm{c}}n_{\mathrm{w}}}\\ \nonumber+ \delta_{m_{\mathrm{h}},n_{\mathrm{h}-1}}\delta_{m_{\mathrm{c}},n_{\mathrm{c}+1}}\delta_{m_{\mathrm{w}},n_{\mathrm{w}+1}}\sqrt{n_{\mathrm{h}}(n_{\mathrm{c}}+1)(n_{\mathrm{w}}+1)}\big)
\end{eqnarray}
and the first-order energy eigenstates are 
\begin{eqnarray}\label{firstorder}
|n^{1}\rangle= \sum_{m\neq n}\frac{\langle m^{(0)}|H_{\mathrm{int}}|n^{(0)}\rangle_{\mu}}{E^{(0)}_{n}-E^{(0)}_{m}}|m^{(0)}\rangle
\end{eqnarray}
where $|m^{(0)}\rangle=\frac{1}{\sqrt{m_{\mathrm{h}}!m_{\mathrm{c}}!m_{\mathrm{w}}!}} z^{m_{\mathrm{h}}}_{\mathrm{h}}z^{m_{\mathrm{c}}}_{\mathrm{c}}z^{m_{\mathrm{w}}}_{\mathrm{w}}$ is the energy eigenstates of index $m=(m_{\mathrm{h}},m_{\mathrm{c}},m_{\mathrm{w}})$ and $E^{(0)}_{n}-E^{(0)}_{m}=\hbar\omega_{\mathrm{int}}(n-m)=\hbar\omega_{\mathrm{int}}(n_{\mathrm{h}}+n_{\mathrm{c}}+n_{\mathrm{w}}-m_{h}-m_{\mathrm{c}}-m_{\mathrm{w}})$. By plugging the expressions of  $\langle m^{(0)}|H_{\mathrm{int}}|n^{(0)}\rangle_{\mu}$ and $E^{(0)}_{n}-E^{(0)}_{m}$ in \ref{firstorder} , we found  the first-order correction of the energy eigenstates to be independent of $\hbar \omega_{\mathrm{int}}$. \\

In the absorption refrigerator , the noise substitutes the work bath and its contact leading to the following interaction term 
\begin{equation}
H_{\mathrm{int}}= g(t) \left(z_{\mathrm{h}} \frac{\partial}{\partial z_{\mathrm{c}}}+ z_{\mathrm{c}} \frac{\partial}{\partial z_{\mathrm{h}}}\right)= g(t) X_{1},
\end{equation}
where $g(t)$ is the stochastic noise field with zero mean $\langle g(t)\rangle=0$ and delta time correlation $\langle g(t)g(t^{\prime})\rangle= 2 \eta \;\delta\left(t-t^{\prime}\right)$
and $X_{1}$ is the generator of swap operation between the two oscillators and belongs to the $\mathrm{SU(2)}$ Lie algebra together with the following operators 
\begin{align}
X_{2}= i \left(z_{\mathrm{h}} \frac{\partial}{\partial z_{\mathrm{c}}}- z_{\mathrm{c}} \frac{\partial}{\partial z_{\mathrm{h}}}\right), \\
X_{3}= \left(z_{\mathrm{h}}\frac{\partial}{\partial z_{\mathrm{h}}}- z_{\mathrm{c}}\frac{\partial}{\partial z_{\mathrm{c}}}\right),\\
N= \left(z_{\mathrm{h}}\frac{\partial}{\partial z_{\mathrm{h}}}+ z_{\mathrm{c}}\frac{\partial}{\partial z_{\mathrm{c}}}\right), 
\end{align}
forming a closed set of equations. The Heisenberg equation for an arbitrary  time-independent analytic function $O(z_{\mathrm{i}}, \overline{z}_{\mathrm{i}})$ reads 
\begin{equation}
\frac{d}{dt }O= -\frac{i}{\hbar}[H_{\mathrm{s}},O]+ \mathcal{L}_{\mathrm{n}}(O)+ \mathcal{L}_{\mathrm{h}}(O)+ \mathcal{L}_{\mathrm{c}}(O), 
\end{equation}
Where the noise dissipator for Gaussian noise is $\mathcal{L}_{\mathrm{n}}(O)= -\eta [X_{1},[X_{1},O]]$ \cite{levy}. For simplicity we assume the heat baths to be uncorrelated between themselves and also uncorrelated with the driving noise. In the Bargmann representation, the hot and cold dissipators are 
\begin{align}
\mathcal{L}_{\mathrm{h}}(O)= \gamma_{\mathrm{h}} \left(\overline{n}_{\mathrm{h}}+1\right)\; \left(z_{\mathrm{h}} O \overline{z}_{\mathrm{h}}-\frac{1}{2}\{z_{\mathrm{h}}\overline{z}_{\mathrm{h}}, O\}\right)\\ \nonumber
+ \gamma_{\mathrm{h}} \overline{n}_{\mathrm{h}} \; \left(\overline{z}_{\mathrm{h}}O z_{\mathrm{h}}- \frac{1}{2}\{\overline{z}_{\mathrm{h}}z_{\mathrm{h}},O\}\right), \\
\mathcal{L}_{\mathrm{c}}\left(O\right)= \gamma_{\mathrm{c}} \left(\overline{n}_{\mathrm{c}}+1\right)\; \left(z_{\mathrm{c}} O \overline{z}_{\mathrm{c}}-\frac{1}{2}\{z_{\mathrm{c}}\overline{z}_{\mathrm{c}}, O\}\right)\\ \nonumber
+ \gamma_{\mathrm{c}} \overline{n}_{\mathrm{c}} \; \left(\overline{z}_{\mathrm{c}}O z_{\mathrm{c}}- \frac{1}{2}\{\overline{z}_{\mathrm{c}}z_{\mathrm{c}},O\}\right),
\end{align}

For a vanishing stochastic driving field, these equations guide separately  the oscillators of  hot and cold baths to thermal equilibrium provided that $\overline{n}_{i}= [e^{-\hbar \omega_{\mathrm{i}}/k_{\mathrm{B}}T_{\mathrm{i}}}-1]^{-1}$ where $\mathrm{i}=\mathrm{h},\mathrm{c}$. To obtain the  cooling current $\mathcal{J}_{\mathrm{c}}= \langle \mathcal{L}_{\mathrm{c}} (\hbar \omega_{c}\; z_{c}\frac{\partial}{\partial z_{c}})\rangle$, we search for the stationary solutions of $X_{3}$ and $N$ we find 
\begin{equation}
\mathcal{J}_{\mathrm{c}}= \hbar \omega_{\mathrm{c}} \frac{\overline{n}_{\mathrm{c}}-\overline{n}_{\mathrm{h}}}{(2\eta)^{-1}+ \gamma_{\mathrm{c}}^{-1}+ \gamma_{\mathrm{h}}^{-1}}. 
\end{equation}
Obviously cooling occurs when $\overline{n}_{\mathrm{c}}-\overline{n}_{\mathrm{h}}>0$. The coefficient of performance $\left(\mathrm{COP}\right)$ is 
\begin{equation}
\mathrm{COP}=\frac{\mathcal{J}_{\mathrm{c}}}{\mathcal{J}_{\mathrm{n}}}= \frac{\omega_{\mathrm{c}}}{\omega_{\mathrm{h}}-\omega_{\mathrm{c}}}\leq \frac{T_{\mathrm{c}}}{T_{\mathrm{c}}-T_{\mathrm{h}}}.
\end{equation}
In this section, we have considered Levy and Kosloff model driven by Gaussian noise for the quantum refrigerator, however it is possible to consider the  Poisson noise case \cite{levy}or more enhanced models presented in \cite{Adesso}. 
\section{The computational complexity}

Bargmann representation of quantum absorption refrigerators and in principle any quantum heat engine has some advantages comparing with the standard treatment based on the operators $a$ and $a^{\dagger}$ that act in   Hilbert space $L^{2}(\mathbb{R}^{d})$. To put this into context we compute the computational complexity of both pictures for a specific example. By definition, the computational complexity is the time consumed by a multitape Turing machine in performing computational tasks\cite{complexity}. The coordinate representation of any  wavefunction in  the bosonic hot or cold reservoir is  given by a Hermite polynomial of degree $n$ multiplied by exponential function  of Gaussian signature up to some numerical constants proportional with the ground-state length scale $\xi= \sqrt{\hbar/2m\omega}$. However in Bargmann representation the wavefunctions of the bosonic hot or cold reservoirs are simply monomials of $z$ with power $n$. As an example consider the ground-state of one specific state in hot or cold reservoir, it has simply the formula $\psi_{0}(x)=( \frac{m \omega}{\pi \hbar})^{1/4} e^{-m\omega x^{2}/2\hbar}$  with computational cost of order  $O(N^{5/2})$ using schoolbook multiplication algorithms and of order $O(N^{2.085})$ using Karatsuba multiplication algorithm where $N$ denotes the input's  digits number. This complexity is simplification of the problem since  we considered the  quantities such as $m,\omega, \hbar,x^{2}$ and other possible  combinations to  be given previous to calculations and thus have complexity of order $O(1)$   \cite{complexity}. However it is very large comparing with the computational cost of the ground-state in Bargmann representation which is of order $O(1)$. The situation becomes very drastically complicated considering higher excited states. 
\vskip 5mm
Another interesting example which shows the advantage of the developed formalism in this work  appears in the calculations of quasiprobability distributions in phase space \cite{Wahiddin,Walls}.  More concretely, the Husimi-Kano $Q$ representation has a simpler form in the Bargmann representation comparing with its form in the phase space $\mathbb{R}^{d}$, and can be written formally in natural units for normalized state $||\psi||=1$  as $H_{\psi}=(2\pi)^{-d} |A\psi|^{2}(\frac{x-ip}{\sqrt{2}}) e^{-|z|^{2}/2}$ \cite{Hall,Husimi}. 
This simplification is mainly because instead of representing $Q$ in term of the conjugate coordinates $X$ and $P$ in the phase space , we simply unify the treatment using the phase space coordinate $z$ only. Thus Bargmann space can be regarded as a natural home for the Husimi-Kano $Q$ representation and other quasi-probability distributions in the phase space\cite{Walls}. In the context of heat engines and refrigerators, the Husimi-Kano $Q$-representation has been used in the calculation of the quantum synchronization since the synchronization measure $S$ is the integral of  $Q$ up to a numerical constant\cite{Armour,Eneriz,Fazio}. Thus it is legitimate to claim that quantum synchronization formalism  simplifies using Bargmann representation and this in turn reduces the computational complexity of the problem. 
\section{Conclusion}
In this work, we discussed the  analytical theory of open quantum systems using  Bargmann representation of the bosonic raising and lowering operators systematically. We also  provided  similar procedure for fermions in term of the anti-commuting Grassmann variables $\left\{\theta_{i}\right\}$ and their partial derivatives. This  construction is useful in two counts first it allows us to exploit the whole theory of analytical functions and all its techniques throughout the computation  of open system characterizations such as the dissipators.  Moreover  it appears to be conceptually  easier to understand than the standard canonical approach based on the raising and lowering  operators $a$ and $a^{\dagger}$ especially for the bosonic case where there is no upper bound on the number of excited states a particle can take. \\ More precisely, we have considered 
the quantum absorption refrigerator driven by a Gaussian noise as a model example. However, the  holomorphic representation is applicable in  all heat engines and refrigerators. We discussed the computational complexity, the time required by a multitape Turing machine to perform specific tasks,  associated with both standard and Bargmann representation for excitations  in bosonic heat baths. We found that working within  Bargmann representation has  less computational complexity  comparing with the standard algebraic or analytical methods in coordinate representation. Another advantage comes from the fact that in holomorphic picture, we use a phase coordinate $z$ instead of the canonical variables $X$ and $P$. This fact simplifies the computation of quasiprobability distributions defined normally in phase space such as the Husimi-Kano $Q$ distributions and this might have impact on the numerical  studies of quantum synchronization in heat engines and in principle for any quantum system. 
\vskip 5mm
{\bf Acknowledgment }\\
We are grateful to  USIM for support. 
MRW acknowledges the support of IIUM research grant IRF19-037-0037.

\end{document}